\documentclass[12pt]{article}
\usepackage{amsmath}
\usepackage{graphicx}
\usepackage{enumerate}
\usepackage{natbib}
\usepackage{url} 

\usepackage{amsmath, amsfonts, bbm, rotating, enumitem}
\usepackage{setspace}
\usepackage{hyperref}
\usepackage{nameref}
\usepackage{verbatim}
\usepackage{natbib}
\usepackage{lmodern,xcolor}
\usepackage{graphicx}
\usepackage{amssymb}
\usepackage{bm}
\usepackage{comment}

\newcommand{\bs}{\bm s}

\newcommand{\blambda}{\bm \lambda}

\newcommand{\bpi}{\bm \pi}
\newcommand{\bphi}{\bm \phi}

\newcommand{\bpsi}{\bm \psi}
\newcommand{\bnu}{\bm \nu}
\newcommand{\bmu}{\bm \mu}

\newcommand{\bxi}{\bm \xi}

\usepackage{caption}
\usepackage{subcaption}
\usepackage{tikz}
\usetikzlibrary{shapes,decorations,arrows,calc,arrows.meta,fit,positioning,shadows}
\tikzset{
    every picture/.style=thick,
    -Latex,auto,node distance =1 cm and 1 cm,semithick,
    state/.style ={ellipse, draw, minimum width = 0.7 cm},
    point/.style = {circle, draw, inner sep=0.04cm,fill,node contents={}},
    bidirected/.style={Latex-Latex,dashed},
    el/.style = {inner sep=2pt, align=left, sloped}
}

\def\bSig\mathbf{\Sigma}

\usepackage{graphicx} 
\usepackage{pifont}

\begin{document}

\title{Flexible Bayesian Tensor Decomposition for Verbal Autopsy Data}

\author{Yu Zhu and Zehang Richard Li$^\ast$\\[4pt]
\textit{Department of Statistics, University of California, Santa Cruz}
\\[2pt]
{lizehang@ucsc.edu}}

\markboth%
{Zhu and Li}
{Tensor Decomposition for Verbal Autopsy Data}

\maketitle

\footnotetext{To whom correspondence should be addressed.}

\begin{abstract}
{
Cause-of-death data is fundamental for understanding population health trends and inequalities as well as designing and evaluating public health interventions. A significant proportion of global deaths, particularly in low- and middle-income countries (LMICs), do not have medically certified causes assigned. In such settings, verbal autopsy (VA) is a widely adopted approach to estimate disease burdens by interviewing caregivers of the deceased. Recently, latent class models have been developed to model the joint distribution of symptoms and perform probabilistic cause-of-death assignment. A large number of latent classes are usually needed in order to characterize the complex dependence among symptoms, making the estimated symptom profiles challenging to summarize and interpret. In this paper, we propose a flexible Bayesian tensor decomposition framework that balances the predictive accuracy of the cause-of-death assignment task and the interpretability of the latent structures. The key to our approach is to partition symptoms into groups and model the joint distributions of group-level symptom sub-profiles. The proposed methods achieve better predictive accuracy than existing VA methods and provide a more parsimonious representation of the symptom distributions. We show our methods provide new insights into the clustering patterns of both symptoms and causes using the PHMRC gold-standard VA dataset. 
}
{Bayesian hierarchical model; probabilistic tensor decomposition; cause-of-death classification; verbal autopsy; mortality quantification.}
\end{abstract}

\section{Introduction}
\label{sec:intro}

Understanding cause-of-death distributions is vital for assessing the health status and needs of a population, particularly in low-resource settings where data scarcity often hampers public health efforts. In many low- and middle-income countries (LMICs), a substantial proportion of deaths occur outside of medical facilities, leaving these deaths unregistered and without medically certified causes of death. To bridge this information gap, verbal autopsy (VA) has emerged as a widely adopted tool, providing an alternative means of cause-of-death determination through structured interviews with a deceased individual’s family or caregiver. VA helps infer causes of death by collecting data on demographics, symptoms, and circumstances preceding death, especially in settings lacking complete civil registration and vital statistics systems \citep{Maher2010health, Sankoh2021epidemiology, Nkengasong2020birth}.

The process of assigning causes of death using VAs traditionally relies on physicians reviewing the collected data and assigning likely causes of death manually. While effective, physician-based VA assessment is resource-intensive, challenging to scale, and often impractical for routine surveillance. As a result, algorithmic and statistical methods have gained traction as cost-effective, scalable alternatives for automated cause-of-death determination. 
Early VA cause-of-death assignment methods usually assume that symptoms are conditionally independent given the underlying cause of death. This simplifying assumption reduces model complexity and computation time, making these algorithms more accessible for large-scale use in low-resource settings \citep{byass2019integrated, Serina2015VA, mccormick2016probabilistic}. However, real-world symptom patterns usually exhibit complex dependence relationships that are crucial for understanding disease risks and accurate cause-of-death assignment. Recent work on verbal autopsy analysis largely focuses on relaxing this assumption. \citet{tsuyoshi2017} and \citet{moran2021bayesian} proposed latent Gaussian factor models to account for symptom dependence and their association with demographic covariates. \citet{li2020using} proposed a latent Gaussian graphical model to characterize latent conditional independence relationship across symptoms. 
One of the main limitations to the latent Gaussian approach to modeling symptom dependence is that the inferred dependence structures cannot be easily interpreted on the scale of the observed discrete response variables. The latent Gaussian models are also computationally more expensive to estimate. \citet{li2021bayesian} and \citet{wu2021tree} addressed these challenges by developing a latent class model framework for VA, where the conditional distribution of binary symptoms given each cause of death is approximated with a finite mixture model. The symptoms are assumed to be conditionally independent given the individual-level latent class membership. The parameters of the latent representation can then be interpreted as latent symptom profiles given each cause of death, a concept known as the symptom-cause-information (SCI) in the VA literature \citep{Clark2018}.

There are two main challenges with latent class analysis introduced in \citet{li2021bayesian} and \citet{wu2021tree}. First, the classical latent class model framework relies on a single latent variable to cluster observations. Deaths assigned to the same cluster are assumed to have conditionally independent and identical symptom distributions. When the assumed number of latent classes is small, the flexibility of the model formulation is limited and the model may not capture nuanced symptom dependence structures for more complex causes of death. On the other hand, models with a large number of latent classes increase the risk of overfitting to the noisy VA data and can degrade the classification performance for the cause-of-death assignment. 
Second, without structural assumptions, standard latent class analysis usually leads to similar latent class profiles on VA data. This phenomenon is known as weak separation \citep{li2023tree} and is common in high-dimensional latent class analysis. The LCVA model proposed in \citet{li2021bayesian} mitigates this issue by assuming the latent symptom profiles share similar elements on most dimensions and differ for only a sparse set of symptoms. However, as we illustrate in Section \ref{sec:model}, when the similarity across latent profiles are induced because of symptom clustering, the sparsity assumption can be ineffective. The weak separation of latent classes makes the characterization and interpretation of high-dimensional symptom profiles difficult.

In this paper, we propose a unified framework using hierarchical tensor decomposition that addresses both limitations. In addition to clustering deaths, we also perform simultaneous symptom clustering. Specicically, we propose two new methods for modeling VA data using group-wise PARAFAC decomposition and collapsed Tucker (c-Tucker) decomposition \citep{Johndrow2017cTucker} Both models partition the high-dimensional symptoms into a small number of groups and construct the full symptom profiles by combining group-specific symptom sub-profiles. Compared to standard latent class models, our models allow us to capture a wider range of dependence structures among symptoms and causes while using a smaller number of parameters. The symptom groups also provide a more interpretable and parsimonious summary of the symptom-cause relationship. We develop a scalable estimation strategy using Markov chain Monte Carlo (MCMC).
Finally, we demonstrate the effectiveness of our model on VA data, showing that the tensor decomposition approach improves both the accuracy of cause-of-death assignments and the interpretability of the latent representations.




The rest of the paper is organized as follows. Section \ref{sec:model} introduces hierarchical tensor decomposition models for VA data. Section \ref{sec:mcmc} describes posterior inference using MCMC. Section \ref{sec:simulation} evaluates the proposed approach with synthetic data and demonstrates the improved performance in cause-of-death assignment from the proposed models compared to existing methods. Section \ref{sec:result} describes an in-depth analysis of a real-world VA dataset using the proposed models. Section \ref{sec:discuss} concludes with a discussion and future work.

\section{Tensor Decomposition for VA Data}\label{sec:model}

Let $X_i \in \{0, 1\}^p$ denote the $p$-dimensional vector of binary signs/symptoms collected by VA and $Y_i \in \{1, ..., C\}$ denote the reference cause of death for the $i$-th death. In this paper, we consider the scenario where we have access to a training dataset where both $X$ and $Y$ are observed and inference is needed for a target dataset where we only observe $X$.  Let $D_i = 1$ indicates the $i$-th death is in the training dataset and $D_i = 0$ the target dataset. The goals of cause-of-death assignment using VA data usually include two components: individual cause-of-death classification for deaths without a reference cause, i.e., predicting $p(Y\mid X)$, and the estimation of cause-specific mortality fractions (CSMF), i.e., $p(Y)$, in the target dataset. We factorize the joint distribution $p(X, Y) = p(Y)p(X \mid Y)$ following the convention VA modeling \citep{KingandLu2008, li2021bayesian}. This factorization corresponds more naturally to the data-generating process, where the majority of signs and symptoms can be treated as consequences of the underlying disease.   It also allows direct estimation of different $p(Y)$ across training and target datasets.

\subsection{Tensor decomposition of conditional symptom distributions}


The key to probabilistic VA models is the modeling of the conditional probability tensor $p(X \mid Y)$. Most existing VA models can be viewed as model-based decomposition of these high-dimensional probability tensors.
Tensor decompositions 
have been successfully applied in various fields, including medical imaging, genomics, social sciences, and recommendation systems.
Two most widely used models in the literature are PARAFAC decomposition \citep{Hitchcock1927TheEO, dunson2009nonparametric} and the Tucker decomposition \citep{Tucker1966}, which we review below in the context of modeling VA symptoms.

\paragraph{PARAFAC decomposition.} 
The PARAFAC decomposition decomposes a tensor into a sum of rank-one tensors, capturing the interactions between multiple modes of the data. The $K$-component element-wise non-negative PARAFAC decomposition of the probability tensor $p(X_{i1} = x_1, ..., X_{ip} = x_p \mid Y_i = c)$ assumes
\begin{align}
 p(X_{i1} = x_1, ..., X_{ip} = x_p \mid Y_i = c) &= \sum_{k=1}^{K} \lambda_{ck} \prod_{j = 1}^{p} \phi_{ckj}^{x_{j}} (1 - \phi_{ckj})^{1 - x_{j}}.
\end{align}

Both $\blambda$ and $\bphi$ are non-negative, and $\sum_{k=1}^{K} \lambda_{ck} = 1$ for $c = 1, ..., C$. The latent class model in \citet{wu2021tree}is a hierarchical version of the probabilistic PARAFAC decomposition with multiple training datasets. The model in \citet{li2021bayesian} uses a sparse PARAFAC decomposition \citep{Zhou2015SparseTensor} by introducing spike-and-slab priors to $\bphi$. Furthermore, when $K = 1$, the decomposition reduces to assuming symptoms are conditionally independent given each cause of death, and it results in the widely used InSilicoVA algorithm \citep{mccormick2016probabilistic}. As $\bphi_c$ characterizes $K$ sets of independent Bernoulli distributions from which the symptoms are generated given the $c$-th cause of death, we refer to $\bphi_c$ as the symptom profiles for the $c$-th cause.


\paragraph{Tucker decomposition.} The Tucker decomposition expresses a tensor as the product of a core tensor and factor matrices along each mode. The core tensor captures the interactions between the modes, while the factor matrices represent the mode-specific transformations. Specifically, we can express the probability tensor $p(X_{i1} = x_1, ..., X_{ip} = x_p \mid Y_i = c)$ in the form of $K$-component Tucker decomposition as below:
\begin{align}
 p(X_{i1} = x_1, ..., X_{ip} = x_p \mid Y_i = c) &= \sum_{k_1=1}^{K} ... \sum_{k_p=1}^{K} \lambda_{ck_1, ..., ck_p} \prod_{j = 1}^{p} \phi_{ck_j, \; j}^{x_{j}} (1 - \phi_{ck_j, \; j})^{1 - x_{j}}.
 \end{align}
Here $\blambda_c$ is a $K^p$-dimensional core probability tensor. The Tucker decomposition provides a more detailed and comprehensive representation of the tensor by capturing mode-specific transformations and inter-mode interactions. However, the interpretability can be more challenging due to the higher dimensionality of the factor matrices and the core tensor. The computational complexity is also significantly increased. Tucker decomposition has not been explored in the context of VA and it is likely not an effective model given the high dimensionality of VA data.


\subsection{Dimension-grouped tensor decomposition}
In order to strike a balance between flexibility, interpretability, and computational efficiency, we note that a salient feature of VA data is that many symptoms/signs are naturally organized in groups.
For instance, for breathing-related issues, VA questionnaires typically ask about a series of related questions, such as whether the person had fast breathing, prolonged trouble breathing, intermittent trouble breathing, increased trouble breathing in certain positions, etc. VA questionnaires also usually include demographic indicators and indicators associated with risk factors such as drinking and smoking. In order to capture this grouping feature among symptoms, we consider two tensor decomposition models: grouped independent PARAFACs and collapsed Tucker (c-Tucker) decomposition. Both formulations were studied in \citet{Johndrow2017cTucker}.

\paragraph{$r$-group independent PARAFACs.} 
Assuming the symptoms can be divided into $r$ groups, where $1 < r < p$, one straightforward extension to the PARAFAC decomposition is to model each group independently with multiple PARAFAC decompositions. We refer to this model as $r$-group independent PARAFACs. Let $s_j \in \{1, ..., r\}$ denote the group membership indicator for the $j$-th symptom. The $r$-group independent PARAFACs model assumes 
\begin{align}
p(X_{i1} = x_1, ..., X_{ip} = x_p \mid Y_i = c) &= \prod_{s = 1}^r  p(\{X_{ij} = x_j : s_j = s\} \mid Y_i = c),
\\
p(\{X_{ij} = x_j : s_j = s\} \mid Y_i = c) &= \sum_{k_s=1}^{K} \lambda_{ck_s} \prod_{j: s_j = s} \phi_{ck_{s_j}, \;j}^{x_{j}} (1 - \phi_{ck_{s_j}, \; j})^{1 - x_{j}}.
 \end{align}

Similar to the PARAFAC decomposition, the dependence of symptoms within the same group can be arbitrarily flexible with a large enough $K$. The symptoms from different groups are assumed to be independent a prior. 

\paragraph{c-Tucker decomposition.} 
The between-group conditional independence assumption is likely too rigid in VA modeling as groups of symptoms and risk factors are likely dependent given the underlying disease. To further introduce dependence among the groups, we adopt the c-Tucker formulation in \citet{Johndrow2017cTucker} and model the mixing weights $\blambda$ using another non-negative PARAFAC decomposition. That is,
\begin{align}
 p(X_{i1} = x_1, ..., X_{ip} = x_p \mid Y_i = c) &= \sum_{k_1=1}^{K} ... \sum_{k_r=1}^{K} \lambda_{ck_1, ..., ck_r} \prod_{j = 1}^{p} \phi_{ck_{s_j}, \;j}^{x_{j}} (1 - \phi_{ck_{s_j}, \; j})^{1 - x_{j}},\\
\lambda_{ck_1, ..., ck_r} &= \sum_{l = 1}^h \nu_{cl} \prod_{s = 1}^r \psi_{clk_s}.
 \end{align}
 
The c-Tucker decomposition reduces to the $r$-group independent PARAFACs when $h = 1$ and the standard PARAFAC when $h = r = 1$. On the other hand, when $r = p$,  it is equivalent to the standard Tucker decomposition. Therefore, it provides an appealing way to control the complexity of the latent structure.

\begin{figure}[!htb]
    \centering
    \includegraphics[width = \textwidth]{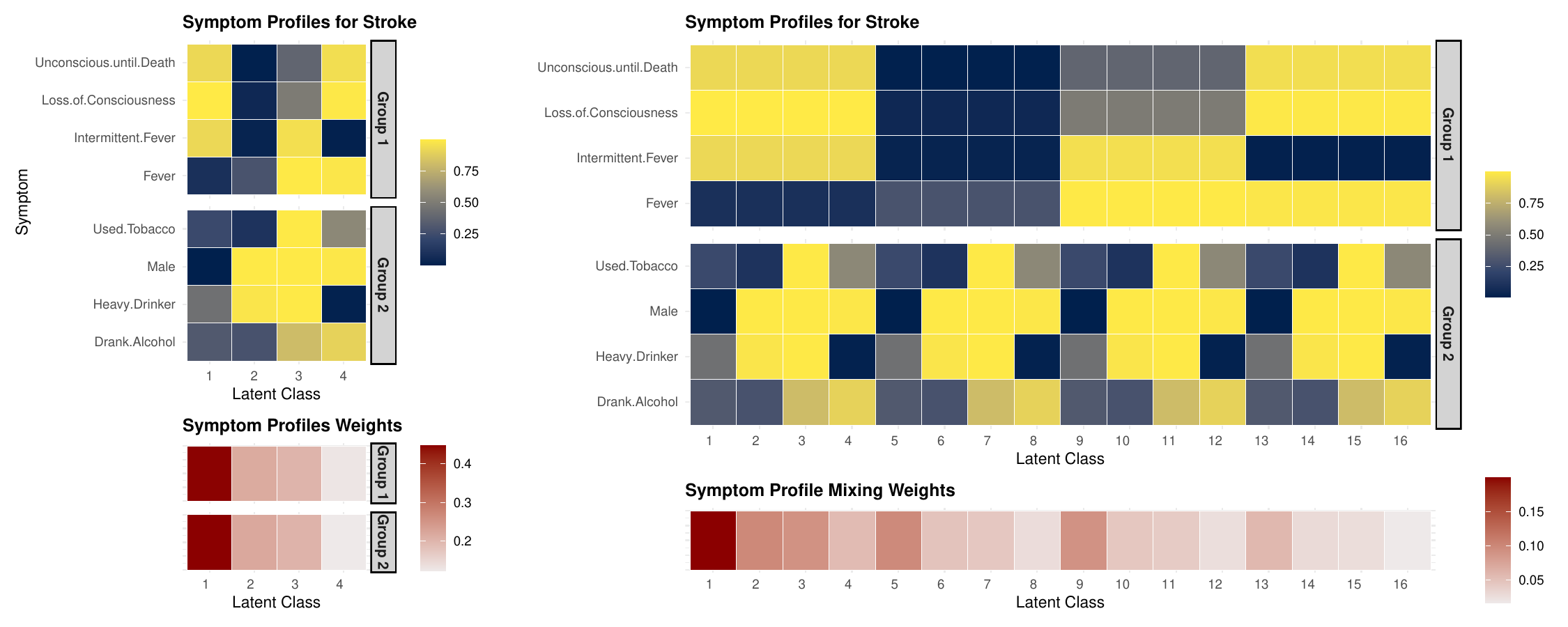}
    \caption{Posterior mean of latent parameters for stroke estimated from the c-Tucker model in one synthetic dataset. Left two panels are the selected symptoms profile $\bm{\phi}_c$ by latent class from 2 symptom groups (top), ordered by the latent class weights $p(Z_{is} = k \mid Y_i = c) = \sum_{l = 1}^h \nu_{cl} \psi_{clsk}$ (bottom). Right two panels are the expanded version of selected symptoms profile $\bm{\phi}$ (top) if the profiles are parameterized under standard PARAFAC decomposition, and the equivalent mixing weights, i.e, $p(Z_{is_1} = k_1, Z_{is_2} = k_2 \mid Y_i = c)$ for $s_1, s_2 \in \{1, 2\}$ and $k_1, k_2 \in \{1, .., 4\}$.}
    \label{fig:cTucker_latent}
\end{figure}

In both dimension-grouped decompositions, the symptom profile $\bphi_c$ is still a $K$ by $p$ matrix, but a different latent class is assigned for each symptom groups. To illustrate the expressiveness of the symptom profiles estimated by the two dimension-grouped factorization models, Figure \ref{fig:cTucker_latent} shows an example of a subset of latent symptom profiles $\bphi$ and the corresponding weights $\blambda$ for deaths due to stroke using the c-Tucker model and the data described in Section \ref{sec:result}. The combinations of these two group-level sub-profiles $\{\phi_{c, k, j}: k \in 1, ..., K, s_j = 1 \}$ and $\{\phi_{c, k, j}: k \in 1, ..., K, s_j = 2 \}$ can be equivalently represented by the standard PARAFAC model with $K^2$ separate latent profiles.
The right panel of Figure \ref{fig:cTucker_latent} illustrates this expanded representation. With $r$ symptom groups and $K$ latent classes, the symptom profiles under the dimension-grouped factorization are parameterized by $pK$ parameters, whereas for the standard PARAFAC is used, it requires $pK^r$ parameters to characterize the same dependence structure, with many repeated elements. In practice, with a finite number of observations, the PARAFAC model will likely not fully utilize $K^r$ latent classes but will approximate it with a smaller rank instead due to prior shrinkage. This leads to a loss of information on the symptom profiles and can make the resulting latent representation more difficult to interpret.

\subsection{Bayesian hierarchical representation}
\begin{figure}[!htb]
    \centering
    \includegraphics[width = \textwidth]{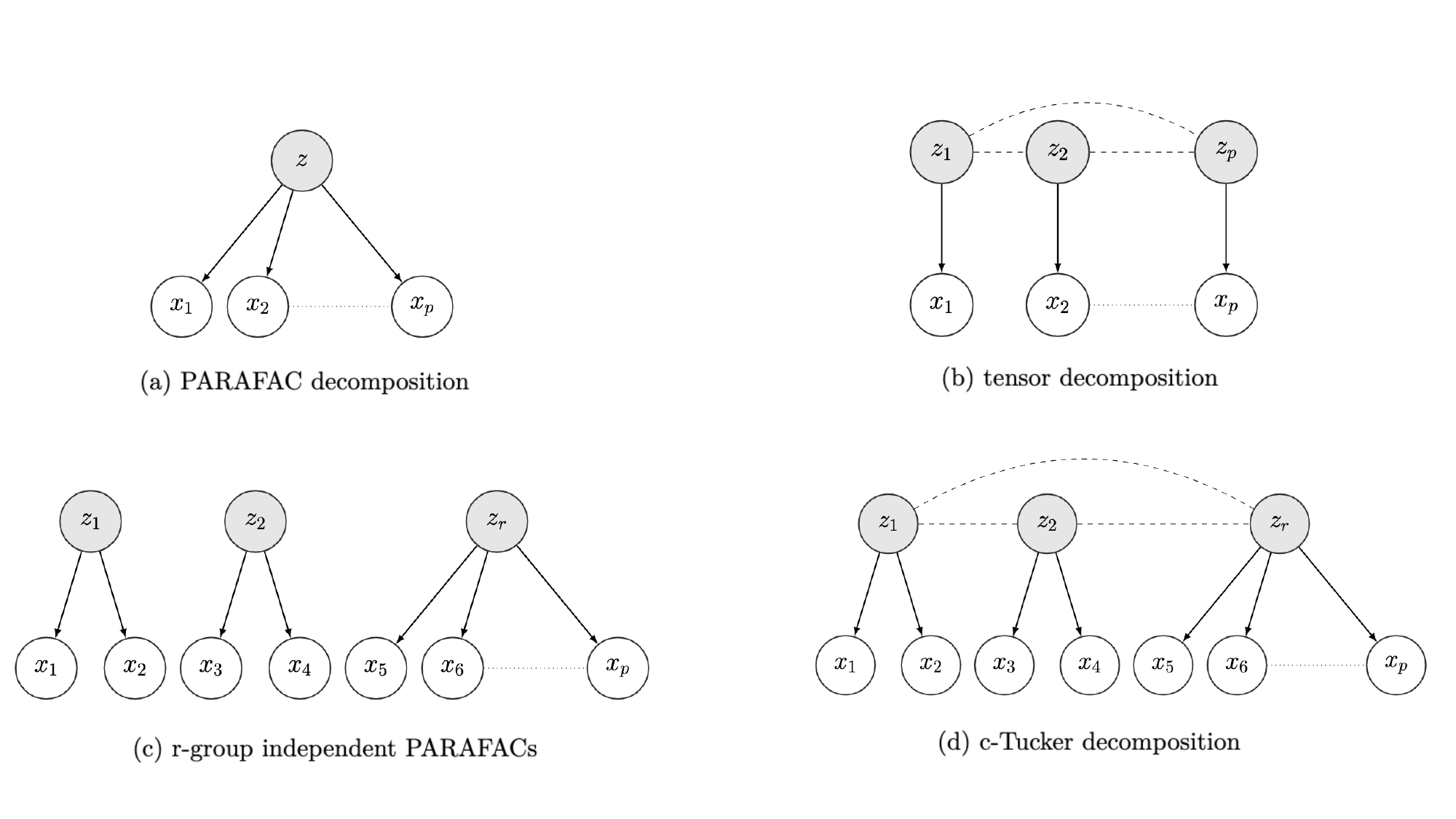}
    \caption{Dependence structures of latent variable $Z$ and indicators $X$ under PARAFAC decomposition, Tucker decomposition, $r$-group independent PARAFACs and c-Tucker decomposition.}
    \label{fig:tensor_decomp}
\end{figure}

To connect the $r$-group independent PARAFACs and c-Tucker decomposition to the latent class models used for cause-of-death assignment, we introduce latent indicators and equivalently represent the conditional probability tensor decomposition using Bayesian hierarchical latent variable models. 
Under both models, we introduce $r$ latent indicators for each death, where $Z_{is} \in \{1, ..., K\}$ specifies latent class membership in the $s$-th symptom group for the $i$-th death. Let $s_{cj} \in \{1, ..., r\}$ denote the group membership for the $j$-th symptom among death due to $c$-th cause. Then the $r$-group independent PARAFACs decomposition can be expressed by  
\begin{align}
Z_{is} \mid Y_i  = c  &\sim Cat(\blambda_{cs}),\\
X_{ij} \mid Y_i  = c, Z_{is_{cj}} = k &\sim Bern(\phi_{ckj}).
\end{align}
For the c-Tucker model, we introduce another latent categorical variable $H_i \in \{1, .., h\}$ to facilitate the dependence among the group-level indicators $Z_{is}$. That is,
\begin{align}
H_i \mid Y_i  = c&\sim Cat(\bnu_c),\\
Z_{is} \mid H_i = l, Y_i  = c  &\sim Cat(\bpsi_{cls}),\\
X_{ij} \mid Y_i  = c, Z_{is_{cj}} = k &\sim Bern(\phi_{ckj}).
\end{align}
A graphical representation for the structure of latent variable $Z$ and indicators $X$, conditional on a single cause of death, under different models is summarized in Figure \ref{fig:tensor_decomp}. 


Finally, we complete the model specification with the priors for $Y$ and the latent parameters. We assume the CSMFs in the training and target datasets can differ, but the conditional distribution of symptoms given causes remain the same, i.e., there is label shift (also known as prior shift) between the joint distributions $p(X, Y)$ in the two datasets \citep{storkey2009training}. This is usually a reasonable assumption when both datasets are collected on the same population. Specifically, for $g \in \{0, 1\}$, we let
\begin{align}
Y_i  \mid D_i = g&\sim \mbox{Cat}(\bpi^{(g)}) \\
\bpi^{(g)} &\sim \mbox{Dir}(\alpha_1, ..., \alpha_C)
\end{align} 

Since we are interested in extracting a small number of groups among symptoms, we use a Dirichlet prior $\blambda_{cs} \sim Dir(a_{s1}, ..., a_{sK})$ under the $r$-group independent PARAFACs. For the c-Tucker model, we let $\bpsi_{cls} \sim Dir(b_{s1}, ..., b_{sK})$, $\bnu_c \sim Dir(1/h, ..., 1/h)$. These Dirichlet priors may be replaced with stick-breaking priors to further encourage sparsity when the latent dimensions are higher. For the response probabilities, we use a non-informative Beta priors, i.e., $\phi_{ckj} \sim Beta(1, 1)$.

As for the symptom grouping, the groups may be predetermined and fixed based on expert knowledge. We adopt a data-adaptive way to estimate the symptom grouping instead by letting
\begin{align}
s_{cj}  &\sim \mbox{Cat}(\bxi_c),\\
\bxi_c &\sim \mbox{Dir}(1/r, ..., 1/r).
\end{align}

\section{Posterior Computation} \label{sec:mcmc}
Posterior distribution of the model parameters can be easily approximated using Gibbs sampling. We describe the MCMC steps for the c-Tucker model first and then the modifications to the $r$-group independent PARAFACs model.

\begin{enumerate}[itemsep=-2mm]
	\item Sample $Y_i \mid \bm{X_i}, \bm{s}, \bpi, \bm{\phi}, \bm{\psi}, \bm{\nu}$ for $i$ where $L_i = 0$ with
   			\[
				p(Y_i = c \mid \bm{X_i}, \bm{s}, \bpi, \bm{\phi}, \bm{\psi}, \bm{\nu}) \propto  
    \pi_c^{(g)} \sum_{k_1 = 1}^K ... \sum_{k_r = 1}^K (\sum_{l =1}^h \nu_{cl} \prod_{s = 1}^r \psi_{clsk_s}) \prod_{j = 1}^p 
    \phi_{ck_{s_{cj}}j}^{X_{ij}}(1-\phi_{ck_{s_{cj}}j})^{1-X_{ij}}
			\]
  \item Sample $Z_{is} \mid H_i = l, Y_i = c, \bm{X_i}, \bm{s}, \bm{\phi}, \bm{\psi}$ for $i = 1, ..., n$, $s = 1, ..., r$ with 
			\[
				p(Z_{is} = k \mid H_i = l, Y_i = c, \bm{X_i}, \bm{s}, \bm{\phi}, \bm{\psi}) \propto \psi_{clsk} \prod_{j:s_{cj} = s}^p\phi_{ckj}^{X_{ij}}(1-\phi_{ckj})^{1-X_{ij}}
			\]	
    \item Sample $H_{i} \mid Y_i = c, Z_{is}, \bm{\psi}, \bm{\nu}$ for $l = 1, ..., h$ with 
			\[
				p(H_i = l\mid Y_i = c, Z_{is}, \bm{\psi, \bm{\nu}}) \propto \nu_{cl} \prod_{s = 1}^r \psi_{cls, Z_{is}}
			\]	
    \item Sample $\bpi^{(g)} \mid Y$ with
    \[
    \bpi^{(g)} \mid Y \sim \mbox{Dir}\left(\alpha_1 + \sum_{i = 1}^{n_g} \mathbbm{1}_{\{Y_i = 1\}}, ..., \alpha_C + \sum_{i = 1}^{n_g} \mathbbm{1}_{\{Y_i = C\}}\right)
    \]
    \item Sample $\bnu_c \mid Y, H$ with
    \[
    \bnu_c \mid Y \sim \mbox{Dir}\left(1/h + \sum_{i = 1}^{n} \mathbbm{1}_{\{Y_i = c, H_i = 1\}}, ..., 1/h + \sum_{i = 1}^{n} \mathbbm{1}_{\{Y_i = c, H_i = h\}}\right)
    \]
    \item Sample $\bpsi_{cls} \mid Y, H, Z$ with
    \[
    \bpsi_{cls} \mid Y \sim \mbox{Dir}\left(b_{s1} + \sum_{i = 1}^{n} \mathbbm{1}_{\{Y_i = c, H_i = l, Z_{is} = 1\}}, ...,  b_{sK} + \sum_{i = 1}^{n} \mathbbm{1}_{\{Y_i = c, H_i = l, Z_{is} = K\}}\right)
    \]
     \item Sample $\phi_{ckj} \mid \bm{Y}, \bm{X}, \bm{Z}, \bm{s}$,  for $c = 1,...,C$, $k = 1, ...,K$, $j = 1, ..., p$,   with 
		\[
		\phi_{ckj} \mid \bm{Y}, \bm{X}, \bm{Z}, \bm{s} \sim \mbox{Beta}\left(a_\phi + \sum_{i=1}^{n} \mathbbm{1}_{\{Y_i=c, Z_{is_{cj}}=k, X_{ij}=1\}}, b_\phi + \sum_{i=1}^{n} \mathbbm{1}_{\{Y_i=c, Z_{is_{cj}}=k, X_{ij}=0\}}\right)
		\]	
      \item Sample $s_{cj} \mid \bm{Y}, \bm{X}, \bm{Z}, \bm{\phi}, \bm{\xi}$  for $j = 1, ..., p$, $s = 1, ..., r$, $c = 1, ..., C$ with
      	\[
				p(s_{cj} = s \mid \bm{Y}, \bm{X}, \bm{Z}, \bm{\phi}, \bm{\xi}) = \frac{\xi_{cs} \prod_{i:Y_i = c}^{n} \phi_{c,Z_{is},j}^{X_{ij}}(1-\phi_{c, Z_{is}, j})^{1-X_{ij}}}{
				\sum_{s = 1}^{r}\xi_{cs} \prod_{i:Y_i = c}^{n} \phi_{c, Z_{is}, j}^{X_{ij}}(1-\phi_{c, Z_{is}, j})^{1-X_{ij}}
				}
			\]	
        \item   Sample $\bm{\xi_c} \mid \bm{s_c}$  for $c = 1, ..., C$ with
            \[
				\bm{\xi_c} \mid \bm{s_c} \sim \mbox{Dir}\left(1/r + \sum_{j = 1}^p \mathbbm{1}_{\{s_{cj} = 1\}}, ..., 1/r+ \sum_{j = 1}^p \mathbbm{1}_{\{s_{cj} = r\}}\right)
			\]	
\end{enumerate}

As for the $r$-group independent PARAFACs, we adjusted the sampling procedures for $Y_i$, $Z_{is}$ and $\blambda_{cs}$ while the sampling procedures for $\bpi^{(g)}$, $\phi_{ckj}$, $s_{cj}$ and $\bm{\xi_c}$ remain consistent with those used in the c-Tucker model. The specific updates are as follows:

\begin{enumerate}[itemsep=-2mm]
	\item Sample $Y_i \mid \bm{X_i}, \bm{s}, \bpi, \bm{\phi}, \bm{\psi}, \bm{\nu}$ for $i$ where $L_i = 0$ with
			\[
				p(Y_i = c \mid \bm{X_i}, \bm{s}, \bpi, \bm{\phi}, \bm{\psi}, \bm{\nu}) \propto  
    \pi_c^{(g)} \prod_{s = 1}^r \sum_{k = 1}^K \lambda_{csk} \prod_{j:s_j = s}^p 
    \phi_{ckj}^{X_{ij}}(1-\phi_{ckj})^{1-X_{ij}}
			\]
  \item Sample $Z_{is} \mid  Y_i = c, \bm{X_i}, \bm{s}, \bm{\phi}, \bm{\lambda}$ for $i = 1, ..., n$, $s = 1, ..., r$ with 
			\[
				p(Z_{is} = k \mid  Y_i = c, \bm{X_i}, \bm{s}, \bm{\phi}, \bm{\lambda}) \propto \lambda_{csk} \prod_{j:s_j = s}^p\phi_{ckj}^{X_{ij}}(1-\phi_{ckj})^{1-X_{ij}}
			\]	
    \item Sample $\blambda_{cs} \mid Y, Z$ with
    \[
    \blambda_{cs} \mid Y \sim \mbox{Dir}\left(b_{s1} + \sum_{i = 1}^{n} \mathbbm{1}_{\{Y_i = c, Z_{is} = 1\}}, ...,  b_{sK} + \sum_{i = 1}^{n} \mathbbm{1}_{\{Y_i = c, Z_{is} = K\}}\right)
    \]
	
\end{enumerate}



\section{Simulation Study}\label{sec:simulation}
We start with a simulation study using synthetic data to evaluate the performance of different tensor decomposition models in terms of cause-of-death assignment. We consider $C = 20$ causes of death, $p = 80$ symptoms, and a total of $3000$ deaths where the training data consist of $2000$ labeled VAs and target data consist of $1000$ unlabeled VAs. For each training-target pair, we generate \( \boldsymbol{\pi}^{(g)} \sim Dir(1, \dots, 1) \) for $g = 0$ and $1$ independently. The symptoms $X$ are generated according to the c-Tucker model with latent dimensions $K = 3$, $r = 5$ and $h = 3$. For each cause $c$ and latent group $l$, we set $\nu_{cl} = 1/h$, ensuring equal contribution across groups.  We construct the $\bs$ matrix where each row consists of $r$ sequential groups arranged in a cyclic pattern:
\[
s_{cj} = \left\lfloor \frac{(j - (c - 1))\mod p}{g} \right\rfloor + 1,
\]
where $g = p/r$, $1 \leq c \leq C$ and $1 \leq j \leq p$. We consider two scenarios of generating the latent class memberships as follows:

\begin{enumerate}
    \item Scenario I: we generate data according to the proposed c-Tucker model. We consider the more realistic situation where the concentrations of latent classes can vary considerably across causes and symptom groups. We let $\bpsi_{cls} \sim Dir(\beta_{cls}\bm 1_{K})$ for each $c \in \{1, ..., C\}$, $l \in \{1, ..., h\}$ and $s \in \{1, ..., r\}$, where the overall concentration parameters $\beta_{cls}$ are independently drawn from a discrete uniform distribution on $[1, 10]$. 

    \item Scenario II: we generate data according to a misspecified c-Tucker model where additional distribution shift exists between the training and target dataset in terms of the symptom distribution conditional on causes.  In this case, denote $p(Z_{is} = k \mid H_i = l, Y_i = c, D_i = g) = \bpsi_{clsk}^{(g)}$. We simulate $\bpsi_{cls}^{(g)} \sim Dir(\beta_{cls}^{(g)}\bm 1_{K})$ and $\beta_{cls}^{(g)}$ are sampled from independently drawn from a discrete uniform distribution on $[1, 10]$. In this scenario, all models are severely misspecified as $p(X \mid Y)$ can be different across the training and target data, depending on the sampled mixing weights $\bpsi$. 
\end{enumerate}

Finally, to mimic the low signal-to-noise ratio in real VA datasets, we allow only a small set of the symptoms to be informative to cause-of-death classification. More specifically, we let
    \begin{align*}
    X_{ij} &\mid Z_{is_{cj}} = k \sim Bern(\phi_{kj}), \quad \phi_{kj} \sim \text{Beta}(1, 1), \;\; \text{if} \; s_{cj} \in \{1, 2\},
    \\
    X_{ij} &\mid Y_i = c, Z_{is_{cj}} = k \sim Bern(\phi_{ckj}), \quad \phi_{ckj} \sim \text{Beta}(1, 1), \;\; \text{if} \; s_{cj} \in \{3, 4, 5\}.
    \end{align*}

In both scenarios, we generate $50$ synthetic datasets and fit the c-Tucker and the $r$-group independent PARAFACs under the same parameter settings as in the data generating process, i.e.,  $K = 3, r = 5$, and $h = 3$. We compare the models with the standard PARAFAC with different numbers of latent classes, $K = 5, 10$, and $15$. As discussed before, the standard PARAFAC correspond to the c-Tucker model with $h = r =1$. We also include comparisons with the LCVA algorithm described in \citet{li2021bayesian} with $K = 10$. LCVA is based on a version of sparse PARAFAC decomposition of $p(X \mid Y)$ and has been shown to outperform other existing VA methods. Since label shift is assumed for the rest of the models in consideration, we only compare with the single-domain variation of LCVA, where $p(Y)$ is treated to be different but $p(X\mid Y)$ is set to be the same across training and target datasets. It should be noted that LCVA implements the cause-of-death assignment in two stages, which is slightly different from the rest of the models in terms of model estimation procedure. For each  dataset, we fit the proposed model and the PARAFAC model by running MCMC for $3000$ iterations, with the first $1000$ iterations discarded as burn-in. For the LCVA, we follow \citet{li2021bayesian} and run three parallel chains of MCMC for $4000$ iterations with the first $1000$ iteration discarded as burn-in.

The models are compared in terms of two widely used metrics in evaluating cause-of-death assignment: the accuracy of the predicted top cause of death and the CSMF accuracy. The CSMF accuracy metric is widely applied to evaluate how closely the estimated population distribution of causes of death matches the true distribution. It is defined as
\begin{align}
\text{CSMF}_{\text{acc}}(\hat{\pi}) &= 1 - \frac{\sum_{c=1}^C |\hat{\pi}_c - \pi_c|}{2\{1 - \min_c \pi_c\}},
\end{align}
where $\pi_c$ denotes the true CSMF from the target dataset. This metric measures the discrepancy between the estimated and actual CSMF, and is scaled to be between 0 and 1, where higher values indicate better performance in estimating prevalence. 

\begin{figure}[!htb]
    \centering
    \includegraphics[width = 0.85\textwidth]{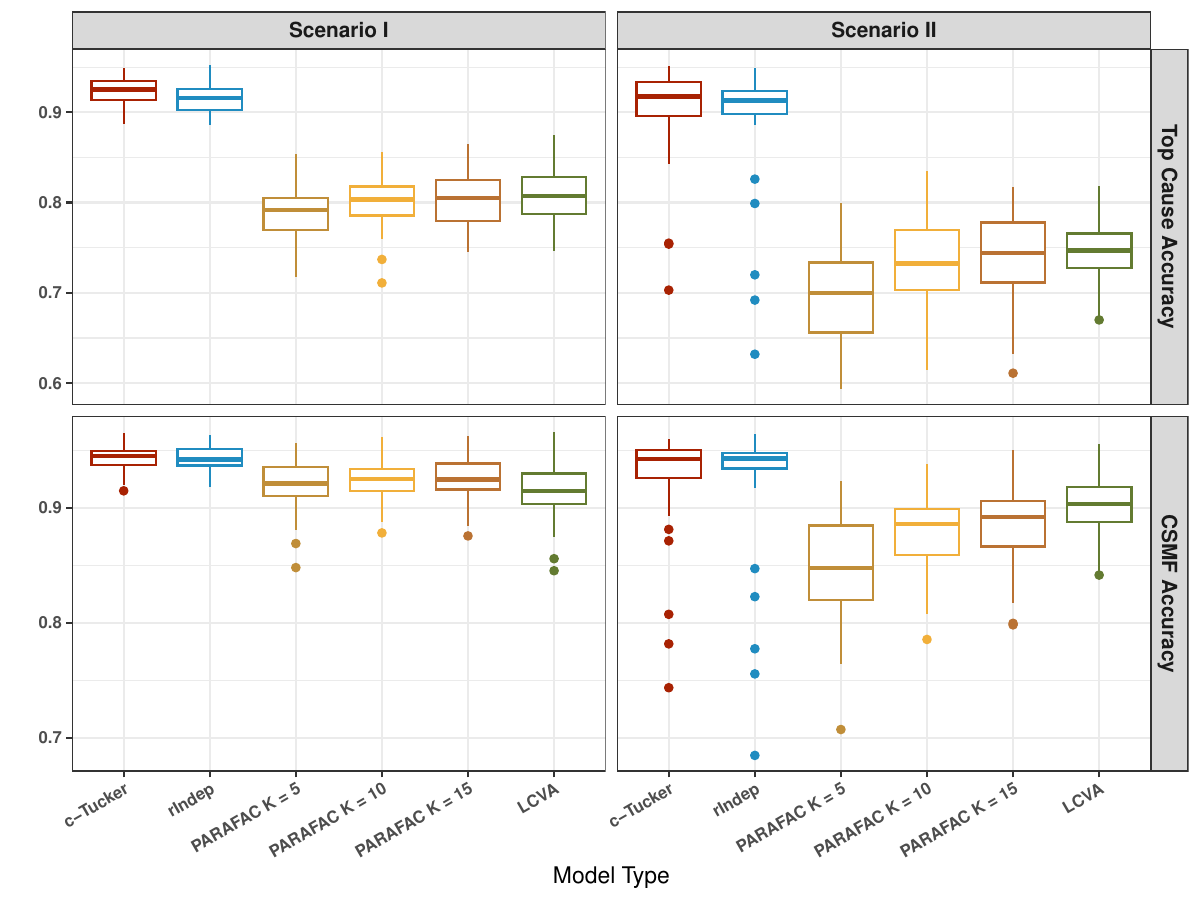}
    \caption{Top cause accuracy (top row) and
CSMF accuracy (bottom row) for the two simulation scenarios, comparing the c-Tucker model, the $r$-group independent PARAFACs model, the PARAFAC with $K = 5, 10, 15$ and LCVA with $K = 10$ on the $50$ simulated datasets.}
    \label{fig:simulation_acc_comp}
\end{figure}

Figure \ref{fig:simulation_acc_comp} summarizes the accuracy measures for different models. Performance of standard PARAFAC generally improve as $K$ gets larger. Both the c-Tucker and $r$-group independent PARAFACs demonstrate higher accuracy compared to all standard PARAFAC models in both scenarios. Furthermore, the c-Tucker and $r$-group independent PARAFACs are more robust to distribution shift across datasets than the PARAFAC and single-domain LCVA, with improvements evident in both the mean and variance of accuracy. While distribution shift is not intentionally accounted for in our model, this scenario shows the pitfalls of more rigid models (e.g., PARAFAC) when they cannot approximate the complex data distribution well.

\section{Analysis of PHMRC Gold-standard Dataset}\label{sec:result}
In this section, we comprehensively evaluate our proposed models using the PHMRC gold-standard VA dataset \citep{murray2011population}. The dataset includes $7841$ adult deaths across 34 distinct causes. This dataset has been processed into $168$ binary symptoms following \citet{mccormick2016probabilistic}. We perform random sampling with replacement from the PHMRC dataset, selecting 80\% of the data as training samples. To ensure that this subset is representative of the entire population, we set the prevalence, $\bpi^{(1)}$, to match the true prevalence observed across the full dataset. We then generate target datasets by first sampling the target prevalence $\bpi^{(0)}$ from $Dir(1, \dots, 1)$ and resampling the remaining observations not in the training data with replacement to match the target prevalence. We repeat the process to form $50$ target datasets and model fittings are carried out in the same way as described in the previous section.

In the rest of this section, we first describe how we choose the number of latent components for the c-Tucker and $r$-group independent PARAFACs models in Section \ref{sec:phmrc-chooseK}. Then we take the c-Tucker model as a case study and further delve into the estimated latent parameters and the interpretation of model components. We discuss the symptom grouping structure learned from data in Section \ref{sec:phmrc-symp} and the implied cause-of-death similarities in Section \ref{sec:phmrc-cause}. Finally, we assess the cause-of-death assignment accuracies of the proposed model and compare them with existing VA methods in Section \ref{sec:phmrc-acc}.

\subsection{Selecting the number of latent components}\label{sec:phmrc-chooseK}
\begin{figure}[!htb]
    \centering
    \includegraphics[width = 0.8\textwidth]{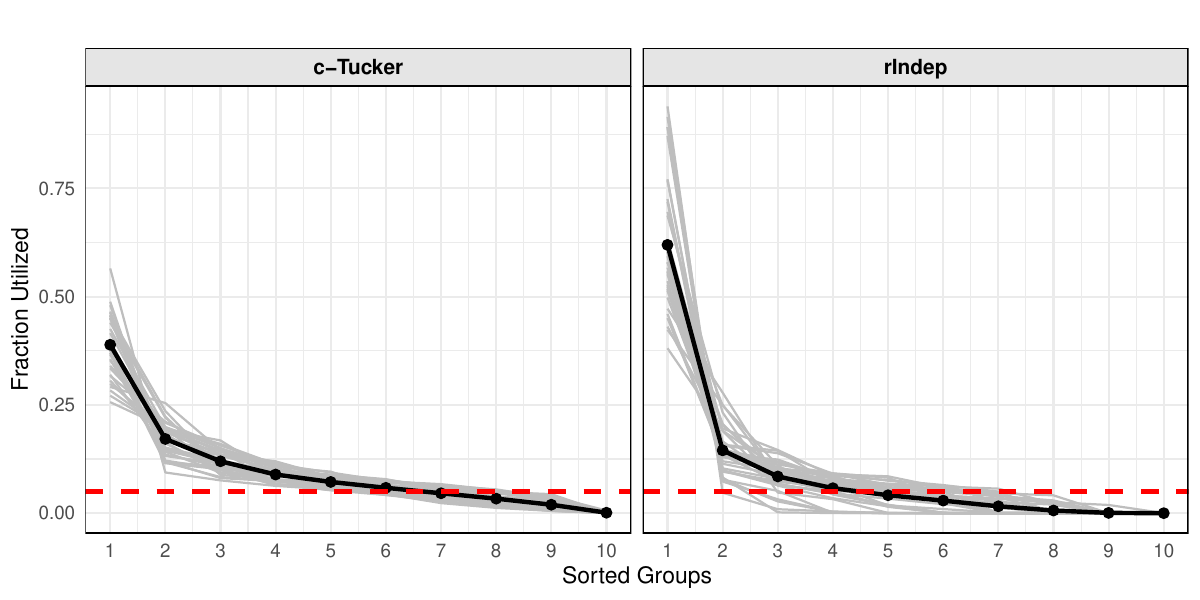}
    \caption{The fraction of times each group is utilized in the posterior samples by the c-Tucker and $r$-group independent PARAFACs models when there are $K = 10$ latent classes. The groups are sorted by the frequency when they are occupied. Each grey line represents one cause of death. The black line represents the simple average across all the causes.}
    \label{fig:rGroup_selection}
\end{figure}

In order to select a reasonable $K$ and $r$,  we first fit both models solely on the training data using large values for $K$ and $r$, setting $K = r = 10$. We select the smallest $r$ that captures at least 95\% of the group variation. Figure \ref{fig:rGroup_selection} illustrates the fraction of times each group is utilized from the posterior samples of $\bs$ for each cause of death, along with average fractions over all causes. For the c-Tucker model, we choose $r = 8$, as the first $8$ groups were utilized in more than $5\%$ of posterior samples. Similarly, we set $r = 6$ for the $r$-group independent PARAFACs model, as the first 6 groups meet this threshold. 

\begin{figure}[!htb]
    \centering
    \includegraphics[width = \textwidth]{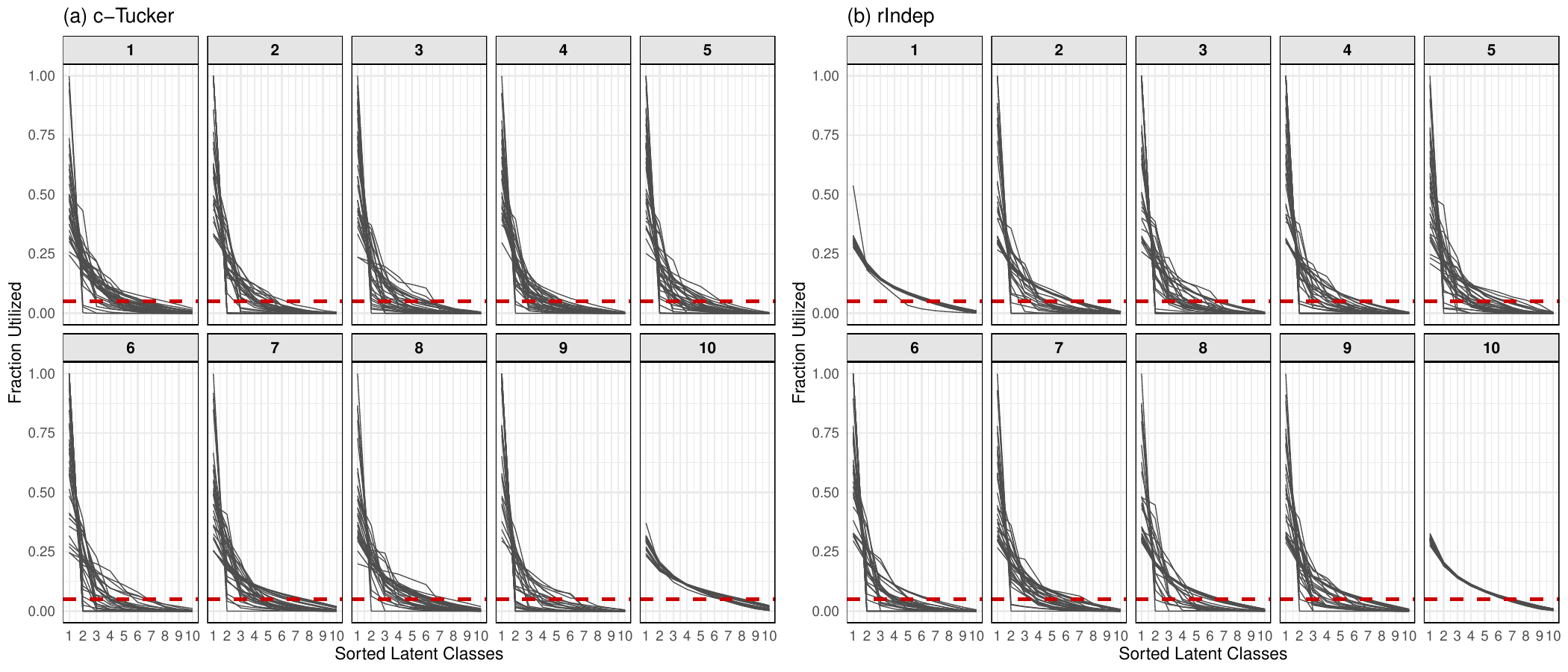}
    \caption{The fraction of times each latent class is utilized in the posterior samples by the c-Tucker and $r$-group independent PARAFACs models when there are $r = 10$ groups. The latent classes are sorted by the frequency when they are occupied in each group. Each green line represents one cause of death.}
    \label{fig:latentK_selection}
\end{figure}

The number of latent classes $K$ is selected similarly. A small $K$ is usually preferrable in practice to avoid overfitting and reduce computational complexity. We choose $K$ so that the majority of latent classes with a high utilization rate can be captured when fitting the models on labeled data. Figure \ref{fig:latentK_selection} shows the proportions of posterior samples where each latent class is utilized, for each symptom group and each cause of death. We let $K = 4$ for the c-Tucker model and $K = 5$ for the $r$-group independent PARAFACs model. This choice is sufficient to capture over $80\%$ of latent classes utilized in at least $5\%$ of posterior samples. More than $99\%$ of such latent classes can be captured by letting $K = 6$ for both models, but in practice, we find larger $K$ usually leads to overfitted models and worse classification performance. As for the c-Tucker model, we set $h = 3$ for the high-level decomposition of the mixing weights since we allow only $8$ symptom groups.
When fitting the standard PARAFAC model, we select $K = 10$ based on the recommendation of \cite{li2021bayesian} on the same dataset.




\subsection{Grouping of symptoms}\label{sec:phmrc-symp}

\begin{figure}[!htb]
    \centering
    \includegraphics[width = \textwidth]{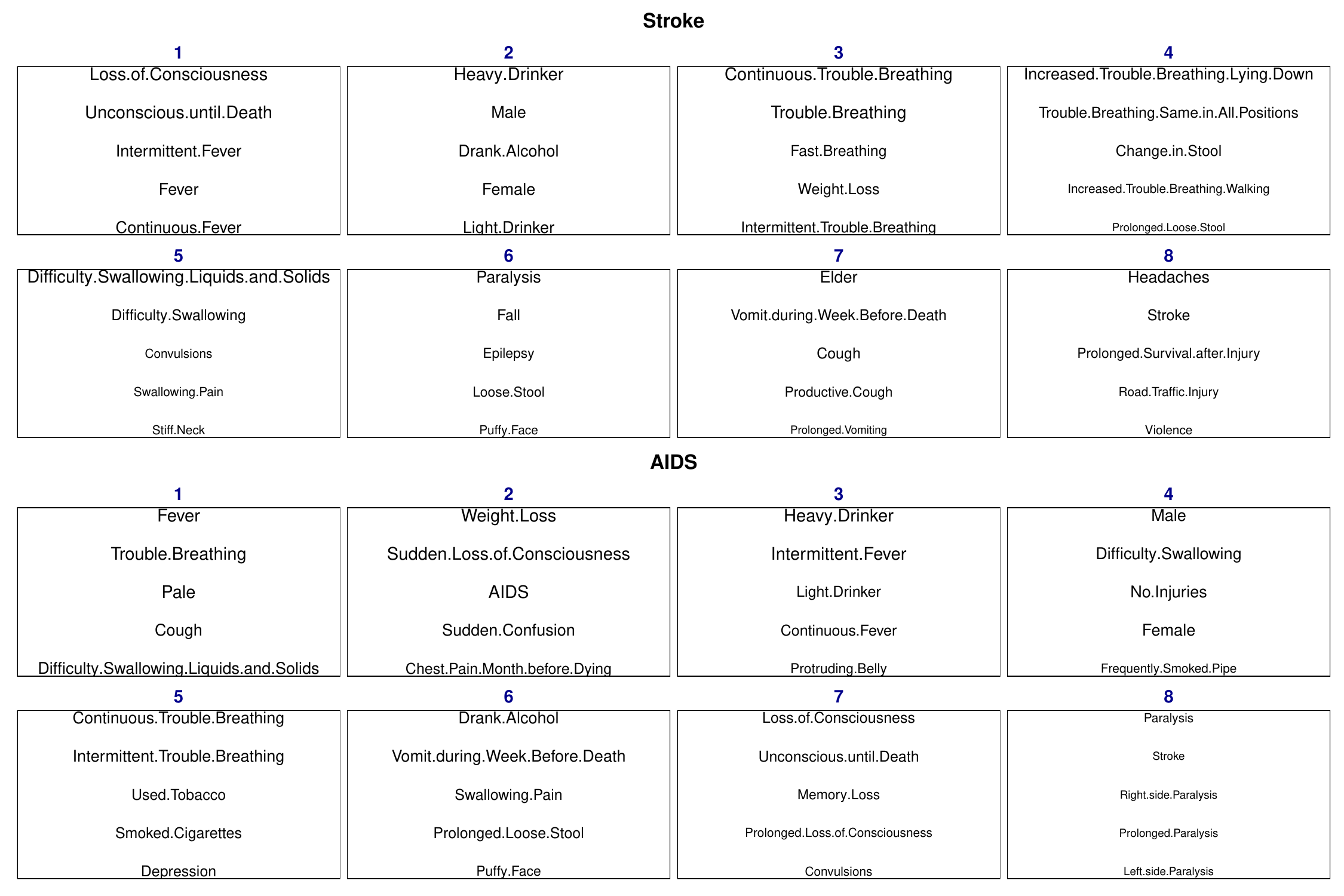}
    \caption{The group structure of key symptoms from the c-Tucker model for two major causes of death: stroke and AIDS. Within each group, symptoms are ordered by their probability of being in that group, listed from highest to lowest. Due to space, only the first five symptoms are shown. For symptoms with the same probability of being in the group, we sort them based on the empirical probability of observing the symptom given the cause, i.e., $p(X_{ij} = 1 \mid Y = c)$. The size of the label is also proportional to this empirical conditional probability. The $8$ groups are re-ordered to have decreasing average empirical conditional probabilities of individual symptoms.}
    \label{fig:symptoms_topic}
\end{figure}

We now illustrate the estimated symptom structures based on the c-Tucker model in one synthetic dataset. Figure \ref{fig:symptoms_topic} presents a summary of key symptom groups for deaths due to two major causes in the PHMRC dataset: stroke and AIDS. We determine the symptom group membership based on the posterior mode of the group indicator matrix $\bs$. 
Symptoms within each group are ordered by the posterior probability of each symptom being within that group, i.e., $p(s_{cj} = \hat s_{cj})$,  in decreasing order. We focus on symptoms with high group assignment probabilities because when this probability is close to $1$, the symptom serves as an anchor to the latent symptom cluster, which has been shown to be key to ensure the identifiability of latent representations \citep{arora2013practical, moran2022identifiable}. Thus, Figure \ref{fig:symptoms_topic} can be viewed as a representation of key symptom ``topics'' within each of the eight groups, analogous to the topic modeling literature. We note that being in the same group does not determine the sign of pairwise correlations. Rather, symptoms in the same group tend to vary together following the $K$ group-level sub-profiles. 


\begin{figure}[!htb]
    \centering
    \includegraphics[width = \textwidth]{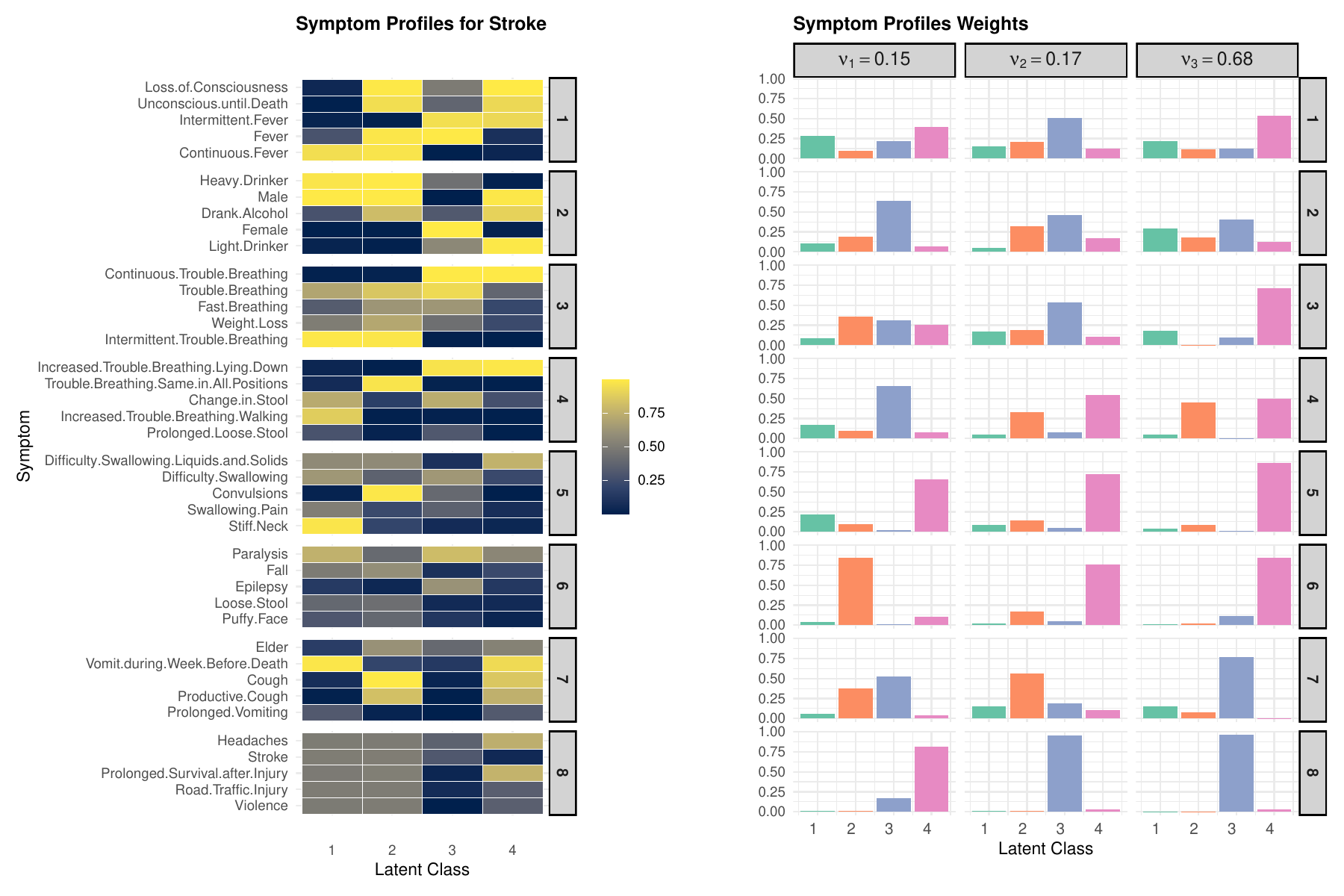}
    \caption{Posterior mean of latent parameters for stroke estimated from the c-Tucker model in one synthetic dataset. The left heatmap is the selected symptoms profile $\bm{\phi}$ by latent class from 8 symptom groups, with the group and symptoms matching the same order as Figure \ref{fig:symptoms_topic}. The right barplots are the symptoms profile latent weights decomposed with $\bm{\psi}$ and corresponding $\bm{\nu}$.}
    \label{fig:cTucker_latent_all}
\end{figure}

We now take a closer look at latent symptom profiles and corresponding mixing weights through the posterior distributions of $\bphi$, $\bmu$, and $\bpsi$. Figure \ref{fig:cTucker_latent_all} illustrates these relationships for the cause of death attributed to stroke estimated by the c-Tucker model. On the left, symptom sub-profiles $\bm{\phi}_c$ are shown for the 8 estimated symptom groups and the first $5$ anchor symptoms within each group, as in Figure \ref{fig:symptoms_topic}. The complete set of symptom profiles can be constructed by mixing the group-level symptom sub-profiles. In this example, $4^8 = 65536$ expanded symptom profiles can be constructed, though many of which have a very low probability of occurrence. 

The relative prevalence of the expanded symptom profiles is determined by the estimated weights in the right panel of Figure \ref{fig:symptoms_topic}. Let $c$ denote the index for stroke,
the weights $\bnu_c$ specifies $h = 3$ set of random weights. In this example, $\nu_{c3} = 0.68$ is the largest latent class on the group weights, indicating that around $68\%$ of deaths due to stroke is expected to follow the third set of group-level latent weights $\bpsi_{c3s}$ for each group $s = 1, ..., 8$. 
While interpreting Figure \ref{fig:cTucker_latent_all} is selective by nature as only a subset of symptoms is shown due to space limitation, we can still observe interesting patterns that match clinical characterizations of stroke deaths.
For this largest cluster, the first symptom group is dominated by latent class $4$, the second symptom group is mostly split between the latent classes $1$ and $3$, etc. Focusing on the symptom profiles with high conditional probabilities, we can characterize this cluster of stroke deaths as having high probability reporting symptoms including loss of consciousness, unconsciousness until death, continuous trouble breathing, and having difficulty swallowing, which are all common comorbidities of severe stroke. They are also expected to split about evenly between having trouble breathing lying down and in all positions, likely depending on the severity of the stroke. The risk of experiencing paralysis is also elevated slightly based on the sixth group of symptoms and the deaths are slightly more likely to be females than males based on the second group of symptoms. In comparison, for example, in the next largest cluster with $\nu_{c2} = 0.17$, the deaths are more likely to report symptoms including fever, cough, productive cough, etc., which may be related to medical complications such as pneumonia, which are also common among stroke deaths.



\subsection{Grouping of causes of death}\label{sec:phmrc-cause}
Estimating cause-specific symptom grouping also allows us to compare causes of death in terms of how symptoms cluster. While we do not assume any information sharing across causes by design, we observe similarities across causes in terms of how symptoms are grouped. For example, comparing stroke and AIDS in Figure \ref{fig:symptoms_topic}, some symptom groups are similar, e.g., symptom clusters related to breathing problems and loss of consciousness, while some other groups are distinct for each specific cause. 

\begin{figure}[!htb]
    \centering
    \includegraphics[width = \textwidth]{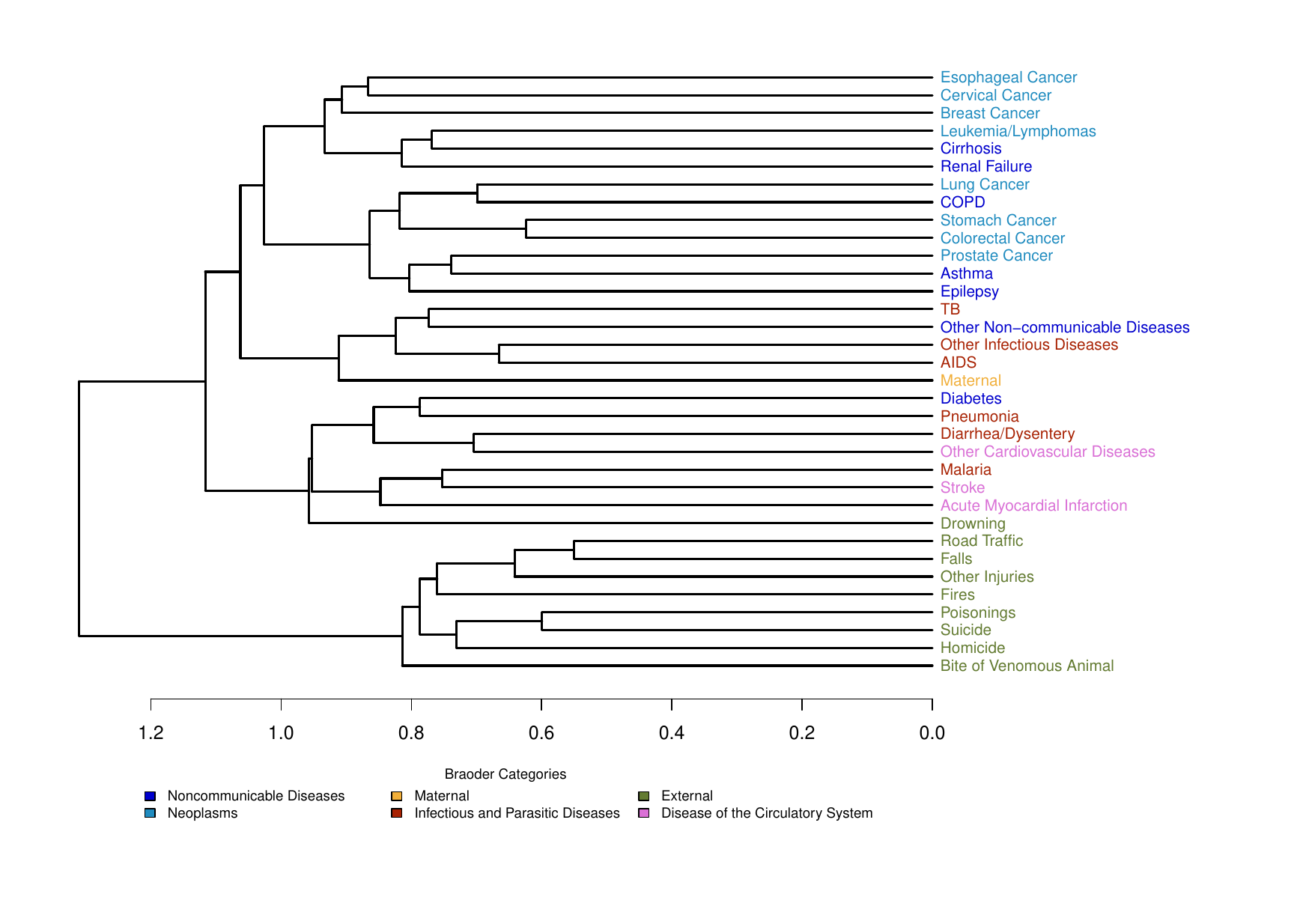}
    \caption{The cause of death dendrogram from the c-Tucker model. The x-axis represents dissimilarity between cause of death labels, calculated as 1 - Adjusted Rand Index (ARI), where ARI measures clustering similarity of symptoms across causes based on the posterior mode of group assignments in $s_{cj}$. The cause of death labels are colored by the 6 broader categories shown in the legend.}
    \label{fig:cod_dendrogram}
\end{figure}

We further explore the similarities of symptom groups across different causes by applying hierarchical clustering to the causes using the estimated $\bs$ matrix.
Figure \ref{fig:cod_dendrogram} shows the dendrogram of all causes of death based on the similarity of $\bs$. The clustering structure reflects dissimilarity between causes, calculated as $1 - \mbox{Adjusted Rand Index (ARI)}$, where ARI  quantifies the similarity in symptom group relationship across different causes of death \citep{Rand1971ARI}. The clustering corresponds very closely to the medically meaningful broader categories of causes. This suggests that the symptom groups discovered by the model are generally similar among related causes.

\subsection{Model performance comparison}\label{sec:phmrc-acc}
To assess the model performance, we compare the results with single-domain LCVA as in the simulation study and InSilicoVA algorithm \citep{mccormick2016probabilistic}. InSilicoVA is a widely used VA method by practitioners. While it has been shown in \citet{li2021bayesian} that LCVA outperforms InSilicoVA in a variety of contexts, we include InSilicoVA as a practical baseline. Other more recent VA algorithms have been shown to perform worse than or comparably to LCVA on the PHMRC dataset and thus we do not include them here for simplicity.

\begin{figure}[!htb]
    \centering
    \includegraphics[width = 0.8\textwidth]{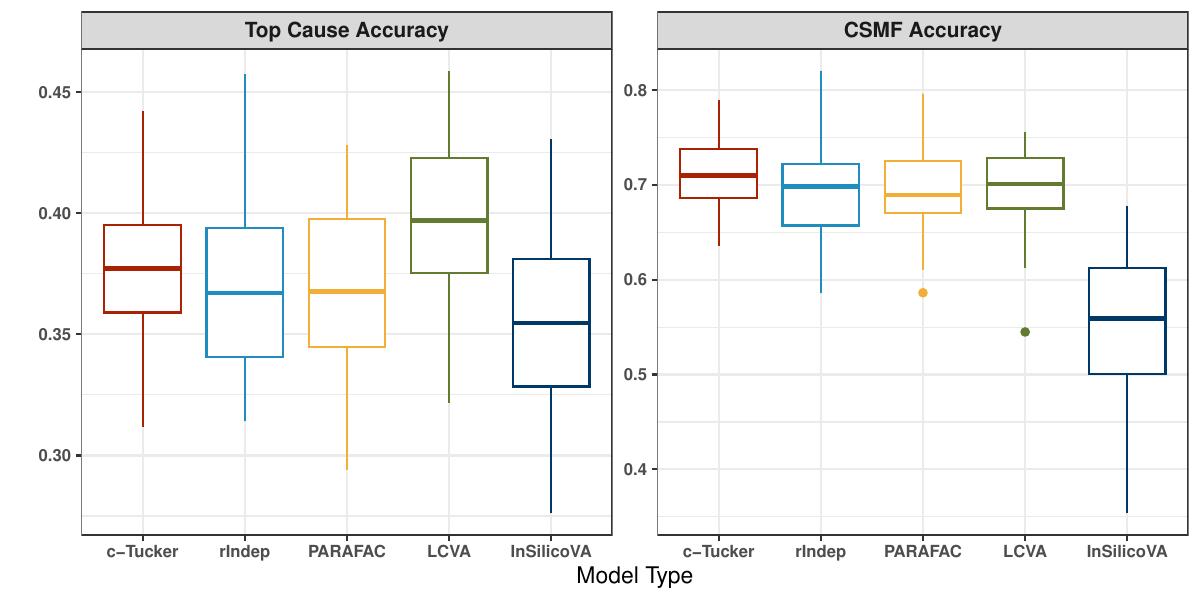}
    \caption{Boxplot of top cause accuracy (left) and CSMF accuracy (right) for different models on the 50 resampled target datasets.}
    \label{fig:phmrc_model_comp}
\end{figure}

Figure \ref{fig:phmrc_model_comp} shows the top cause accuracy and CSMF accuracy across different models on $50$ resampled target datasets. For both metrics, the simplest model, InSilicoVA, exhibits the lowest overall accuracy and the highest variance. The c-Tucker model shows slightly lower top cause accuracy than LCVA but outperforms all methods in CSMF accuracy. We caution against overinterpreting the relative performance of methods on the PHMRC dataset, as it is a single dataset with limited sample size and known issues \citep{byass2014usefulness}. Nevertheless, the numerical analysis shows that the c-Tucker model achieves comparable performance to existing methods and has the potential to be used as a routine analysis method for VAs. 

\section{Discussion}
\label{sec:discuss}

In this paper, we examined the task of modeling VA data through the lens of conditional tensor decomposition. We proposed two dimension-grouped tensor decomposition models for VA data, the collapsed Tucker decomposition and group independent PARAFACs. Both models can identify multiple latent symptom groups and their corresponding profiles, offering a richer and more robust alternative to models based on standard latent class model formulation. Our approach not only captures complex and meaningful dependence structures among symptoms, but also enhances interpretability of the latent parameters by utilizing a more parsimonious parameterization to represent the symptom distributions. Our models also lead to robust improvements in cause-of-death assignment performance at both individual and population levels.

There are several limitations of the proposed model. First of all, while the symptom group structures show strong similarities across related causes of death, we did not explicitly encourage such structure in our model. We do not expect such information sharing to significantly improve predictive performance when the cause-specific sample sizes are moderate to large, but it may be useful to consider with rare diseases and unbalanced datasets. Second, in this paper we only focus on the case of label shift between training and target datasets.The proposed methods can be naturally extended to deal with distribution shift in $p(X\mid Y)$ in a multi-source domain adaptation setting discussed in \citet{li2021bayesian}.
Third, while our model provides a more flexible characterization of the symptom dependence, it comes at the cost of a heavier computational burden. While computation is feasible on a single laptop for the analysis considered in this paper, it can be challenging when scaling up to very large datasets from national mortality surveillance systems. More work needs to be done to enable distributed modeling and inference for big data. Lastly, the latent group structures we consider in this paper may be further condensed into lower-dimensional summaries by adding more model components such as sparsity-inducing priors. It is an important future direction of research to gain more insight into the interpretability and expandability of VA algorithms, especially as more and more black-box classification tools are available in the era of artificial intelligence. We leave these challenges to future work.

\section*{Acknowledgments}

ZY and ZRL were supported by grant R03HD110962 from the Eunice Kennedy Shriver National Institute of Child Health and Human Development (NICHD), and in part by the Bill \& Melinda Gates Foundation. The findings and conclusions contained within are those of the authors and do not necessarily reflect positions or policies of the Bill \& Melinda Gates Foundation. The authors would like to thank Dr. Tsuyoshi Kunihama and Dr. Zhenke Wu for helpful discussion and suggestions.

{\it Conflict of Interest}: None declared.



\bibliographystyle{biorefs}
\bibliography{crms}

\begin{thebibliography}{99}

\bibitem[Arora \emph{and others}(2013)Arora, Ge, Halpern, Mimno, Moitra, Sontag, Wu and Zhu]{arora2013practical}
\textsc{Arora, Sanjeev, Ge, Rong, Halpern, Yonatan, Mimno, David, Moitra, Ankur, Sontag, David, Wu, Yichen and Zhu, Michael}. (2013).
\newblock A practical algorithm for topic modeling with provable guarantees.
\newblock In:  {\em International conference on machine learning\/}. PMLR. pp.\  280--288.

\bibitem[Byass(2014)Byass]{byass2014usefulness}
\textsc{Byass, Peter}. (2014).
\newblock Usefulness of the population health metrics research consortium gold standard verbal autopsy data for general verbal autopsy methods.
\newblock {\em BMC medicine\/}~\textbf{12}, 1--10.

\bibitem[Byass \emph{and others}(2019)Byass, Hussain-Alkhateeb, D’Ambruoso, Clark, Davies, Fottrell, Bird, Kabudula, Tollman, Kahn  et~al.]{byass2019integrated}
\textsc{Byass, Peter, Hussain-Alkhateeb, Laith, D’Ambruoso, Lucia, Clark, Samuel, Davies, Justine, Fottrell, Edward, Bird, Jon, Kabudula, Chodziwadziwa, Tollman, Stephen, Kahn, Kathleen  \emph{and others}}. (2019).
\newblock An integrated approach to processing who-2016 verbal autopsy data: the interva-5 model.
\newblock {\em BMC Medicine\/}~\textbf{17}(1), 1--12.

\bibitem[Clark \emph{and others}(2018)Clark, Li and McCormick]{Clark2018}
\textsc{Clark, Samuel~J, Li, Zehang~R and McCormick, Tyler~H}. (2018).
\newblock Quantifying the contributions of training data and algorithm logic to the performance of automated cause-assignment algorithms for verbal autopsy.
\newblock {\em arXiv: 1803.07141\/}.

\bibitem[Dunson and Xing(2009)Dunson and Xing]{dunson2009nonparametric}
\textsc{Dunson, David~B and Xing, Chuanhua}. (2009).
\newblock Nonparametric {Bayes} modeling of multivariate categorical data.
\newblock {\em Journal of the American Statistical Association\/}~\textbf{104}(487), 1042--1051.

\bibitem[Hitchcock(1927)Hitchcock]{Hitchcock1927TheEO}
\textsc{Hitchcock, Frank~Lauren}. (1927).
\newblock The expression of a tensor or a polyadic as a sum of products.
\newblock {\em Journal of Mathematics and Physics\/}~\textbf{6}, 164--189.

\bibitem[Johndrow \emph{and others}(2017)Johndrow, Bhattacharya and Dunson]{Johndrow2017cTucker}
\textsc{Johndrow, James~E., Bhattacharya, Anirban and Dunson, David~B.} (2017).
\newblock {Tensor decompositions and sparse log-linear models}.
\newblock {\em The Annals of Statistics\/}~\textbf{45}(1), 1 -- 38.

\bibitem[King and Lu(2008)King and Lu]{KingandLu2008}
\textsc{King, Gary and Lu, Ying}. (2008, 09).
\newblock Verbal autopsy methods with multiple causes of death.
\newblock {\em Statistical Science\/}~\textbf{23}.

\bibitem[Kunihama \emph{and others}(2020)Kunihama, Li, Clark and McCormick]{tsuyoshi2017}
\textsc{Kunihama, Tsuyoshi, Li, Zehang~R, Clark, Samuel~J and McCormick, Tyler~H}. (2020).
\newblock {Bayesian factor models for probabilistic cause of death assessment with verbal autopsies}.
\newblock {\em The Annals of Applied Statistics\/}.

\bibitem[Li \emph{and others}(2023)Li, Stephenson and Wu]{li2023tree}
\textsc{Li, Mengbing, Stephenson, Briana and Wu, Zhenke}. (2023).
\newblock Tree-regularized bayesian latent class analysis for improving weakly separated dietary pattern subtyping in small-sized subpopulations.
\newblock {\em arXiv preprint arXiv:2306.04700\/}.

\bibitem[Li \emph{and others}(2020)Li, McComick and Clark]{li2020using}
\textsc{Li, Zehang~Richard, McComick, Tyler~H. and Clark, Samuel~J.} (2020, 09).
\newblock Using {Bayesian} latent {G}aussian graphical models to infer symptom associations in verbal autopsies.
\newblock {\em Bayesian Analysis\/}~\textbf{15}(3), 781--807.

\bibitem[Li \emph{and others}(2024)Li, Wu, Chen and Clark]{li2021bayesian}
\textsc{Li, Zehang~Richard, Wu, Zhenke, Chen, Irena and Clark, Samuel~J}. (2024).
\newblock {Bayesian} nested latent class models for cause-of-death assignment using verbal autopsies across multiple domains.
\newblock {\em Annals of Applied Staitstics\/}~\textbf{18}(2), 1137--1159.

\bibitem[Maher \emph{and others}(2010)Maher, Biraro, Hosegood, Isingo, Lutalo, Mushati, Ngwira, Nyirenda, Todd and Zaba]{Maher2010health}
\textsc{Maher, D., Biraro, S., Hosegood, Victoria, Isingo, R., Lutalo, Tom, Mushati, P., Ngwira, Bagrey, Nyirenda, Makandwe, Todd, Jim and Zaba, Basia}. (2010, 03).
\newblock Translating global health research aims into action: The example of the {ALPHA} network.
\newblock ~\textbf{15}, 321 -- 328.

\bibitem[McCormick \emph{and others}(2016)McCormick, Li, Calvert, Crampin, Kahn and Clark]{mccormick2016probabilistic}
\textsc{McCormick, Tyler~H, Li, Zehang~R, Calvert, Clara, Crampin, Amelia~C, Kahn, Kathleen and Clark, Samuel~J}. (2016).
\newblock Probabilistic cause-of-death assignment using verbal autopsies.
\newblock {\em Journal of the American Statistical Association\/}~\textbf{111}(515), 1036--1049.

\bibitem[Moran \emph{and others}(2022)Moran, Sridhar, Wang and Blei]{moran2022identifiable}
\textsc{Moran, Gemma~Elyse, Sridhar, Dhanya, Wang, Yixin and Blei, David}. (2022).
\newblock Identifiable deep generative models via sparse decoding.
\newblock {\em Transactions on Machine Learning Research\/}.

\bibitem[Moran \emph{and others}(2021)Moran, Turner, Dunson and Herring]{moran2021bayesian}
\textsc{Moran, Kelly~R, Turner, Elizabeth~L, Dunson, David and Herring, Amy~H}. (2021).
\newblock Bayesian hierarchical factor regression models to infer cause of death from verbal autopsy data.
\newblock {\em Journal of the Royal Statistical Society: Series C (Applied Statistics)\/}.

\bibitem[Murray \emph{and others}(2011)Murray, Lopez, Black, Ahuja, Ali, Baqui, Dandona, Dantzer, Das, Dhingra  et~al.]{murray2011population}
\textsc{Murray, Christopher~JL, Lopez, Alan~D, Black, Robert, Ahuja, Ramesh, Ali, Said~M, Baqui, Abdullah, Dandona, Lalit, Dantzer, Emily, Das, Vinita, Dhingra, Usha  \emph{and others}}. (2011).
\newblock Population health metrics research consortium gold standard verbal autopsy validation study: design, implementation, and development of analysis datasets.
\newblock {\em Population Health Metrics\/}~\textbf{9}(1), 27.

\bibitem[Nkengasong \emph{and others}(2020)Nkengasong, Gudo, Macicame, Maunze, Amouzou, Banke, Dowell and Jani]{Nkengasong2020birth}
\textsc{Nkengasong, John, Gudo, Eduardo, Macicame, Ivalda, Maunze, Xadreque, Amouzou, Agbessi, Banke, Kathryn, Dowell, Scott and Jani, Ilesh}. (2020, 01).
\newblock Improving birth and death data for {A}frican decision making.
\newblock {\em The Lancet Global Health\/}~\textbf{8}, e35--e36.

\bibitem[Rand(1971)Rand]{Rand1971ARI}
\textsc{Rand, William~M.} (1971).
\newblock Objective criteria for the evaluation of clustering methods.
\newblock {\em Journal of the American Statistical Association\/}~\textbf{66}(336), 846--850.

\bibitem[Sankoh and Byass(2012)Sankoh and Byass]{Sankoh2021epidemiology}
\textsc{Sankoh, Osman and Byass, Peter}. (2012, 06).
\newblock The indepth network: Filling vital gaps in global epidemiology.
\newblock {\em International Journal of Epidemiology\/}~\textbf{41}, 579--88.

\bibitem[Serina \emph{and others}(2015)Serina, Riley, Stewart, James, Flaxman, Lozano, Hernández-Prado, Mooney, Luning, Black, Ahuja, Alam, Alam, Ali, Atkinson, Baqui, Chowdhury, Dandona and Lopez]{Serina2015VA}
\textsc{Serina, Peter, Riley, Ian, Stewart, Andrea, James, Spencer, Flaxman, Abraham, Lozano, Rafael, Hernández-Prado, Bernardo, Mooney, Meghan, Luning, Richard, Black, Robert, Ahuja, Ramesh, Alam, Nurul, Alam, Sayed~Saidul, Ali, Said, Atkinson, Charles, Baqui, Abdullah, Chowdhury, Dr.~Hafizur, Dandona, Rakhi} \emph{and others}. (2015, 12).
\newblock Improving performance of the tariff method for assigning causes of death to verbal autopsies.
\newblock {\em BMC Medicine\/}~\textbf{13}.

\bibitem[Storkey(2009)Storkey]{storkey2009training}
\textsc{Storkey, Amos}. (2009).
\newblock When training and test sets are different: characterizing learning transfer.
\newblock {\em Dataset shift in machine learning\/}~\textbf{30}, 3--28.

\bibitem[Tucker(1966)Tucker]{Tucker1966}
\textsc{Tucker, Ledyard}. (1966).
\newblock Some mathematical notes on three-mode factor analysis.
\newblock {\em Psychometrika\/}~\textbf{31}(3), 279--311.

\bibitem[Wu \emph{and others}(2024)Wu, Li, Chen and Li]{wu2021tree}
\textsc{Wu, Zhenke, Li, Zehang~Richard, Chen, Irena and Li, Mengbing}. (2024, 02).
\newblock Tree-informed {B}ayesian multi-source domain adaptation: cross-population probabilistic cause-of-death assignment using verbal autopsy.
\newblock {\em Biostatistics\/}, kxae005.

\bibitem[Zhou \emph{and others}(2015)Zhou, Bhattacharya, Herring and Dunson]{Zhou2015SparseTensor}
\textsc{Zhou, Jing, Bhattacharya, Anirban, Herring, Amy~H. and Dunson, David~B.} (2015).
\newblock Bayesian factorizations of big sparse tensors.
\newblock {\em Journal of the American Statistical Association\/}~\textbf{110}(512), 1562--1576.
\newblock PMID: 31210707.

\end{thebibliography}

\end{document}